\documentclass{article}

 \usepackage[dblblindworkshop, final]{neurips_2025}
\workshoptitle{Machine Learning and the Physical Sciences}

\usepackage[utf8]{inputenc} 
\usepackage[T1]{fontenc}    
\usepackage{hyperref}       
\usepackage{url}            
\usepackage{booktabs}       
\usepackage{amsfonts}       
\usepackage{nicefrac}       
\usepackage{microtype}      
\usepackage{xcolor}         
\usepackage{acronym}
\usepackage{defs}
\usepackage{amsmath}
\usepackage{graphicx} 
\usepackage{comment}

\title{Locality-Sensitive Hashing-Based Efficient Point Transformer for Charged Particle Reconstruction}

\author{%
  Shitij Govil\textsuperscript{1,}\thanks{Equal contribution. \quad ${}^\dagger$Corresponding authors: \texttt{ytchou@uw.edu}, \texttt{siqi.miao@gatech.edu}.} \And
  Jack P.~Rodgers\textsuperscript{2,*} \And
  Yuan\mbox{-}Tang Chou\textsuperscript{3,†} \And
  Siqi Miao\textsuperscript{1,†} \And
  Amit Saha\textsuperscript{1} \And
  Advaith Anand\textsuperscript{3} \And
  Kilian Lieret\textsuperscript{4} \And
  Gage DeZoort\textsuperscript{4} \And
  Mia Liu\textsuperscript{2} \And
  Javier Duarte\textsuperscript{5} \qquad
  Pan Li\textsuperscript{1} \qquad
  Shih-Chieh Hsu\textsuperscript{3} \\
  \\
  \textsuperscript{1}Georgia Institute of Technology \quad
  \textsuperscript{2}Purdue University \quad
  \textsuperscript{3}University of Washington \\
  \textsuperscript{4}Princeton University \quad
  \textsuperscript{5}University of California San Diego
}

\begin{document}

\maketitle

\begin{abstract}
Charged particle track reconstruction is a foundational task in collider experiments and the main computational bottleneck in particle reconstruction.
Graph neural networks (GNNs) have shown strong performance for this problem, but costly graph construction, irregular computations, and random memory access patterns substantially limit their throughput.
The recently proposed Hashing-based Efficient Point Transformer (HEPT) offers a theoretically guaranteed near-linear complexity for large point cloud processing via locality-sensitive hashing (LSH) in attention computations; however, its evaluations have largely focused on embedding quality, and the object condensation pipeline on which HEPT relies requires a post-hoc clustering step (e.g., DBScan) that can dominate runtime.
In this work, we make two contributions. First, we present a unified, fair evaluation of physics tracking performance for HEPT and a representative GNN-based pipeline under the same dataset and metrics.
Second, we introduce HEPTv2 by extending HEPT with a lightweight decoder that eliminates the clustering stage and directly predicts track assignments. This modification preserves HEPT’s regular, hardware-friendly computations while enabling ultra-fast end-to-end inference.
On the TrackML dataset, optimized HEPTv2 achieves approximately 28 ms per event on an A100 while maintaining competitive tracking efficiency. These results position HEPTv2 as a practical, scalable alternative to GNN-based pipelines for fast tracking.
\end{abstract}

\section{Introduction}
With many fundamental questions unanswered in particle physics, the \ac{LHC}~\citep{LyndonEvans_2008} at CERN remains the flagship machine for understanding unsolved questions at the energy frontier. The \ac{LHC} is scheduled for significant upgrades in its accelerator complex in the coming year, substantially increasing the number of simultaneous proton-proton interactions in each collision record (pile-up) for experiments such as ATLAS~\citep{ATLAS:2023dns} and CMS~\citep{CMS:2008xjf}, from <$\mu$> = 60 to <$\mu$> = 200. Consequently, a large number of particles need to be reconstructed and processed from the detector measurements. Particle trajectory reconstruction is the most computationally expensive, but key component of particle reconstruction and identification. Traditional rule-based algorithms such as the Combinatorial \ac{KF} have been used primarily for this task~\citep{ATLAS:2024rnw}. However, the unprecedented data rates and complexity of \ac{HL-LHC} conditions demand innovative approaches with lower inference latency.

The development of \ac{ML}-based algorithms has emerged as an alternative solution to tackle these challenges. GNNs have been widely adopted, primarily because they exploit the local inductive bias inherent in particle interactions. However, their expensive graph construction and irregular neighborhood aggregation lead to hardware inefficiencies \citep{Stark2022,MoritzLieret2023,Jang2010,Hashemi2018,Abadal2021}, making GNNs less suitable for the data throughput needed for \ac{HL-LHC}~\citep{CMSGroup2022}.
In another line of work, multiple efficient point transformers have been proposed for large point cloud processing~\citep{Zhao2021PointTransformer,Liu2023FlatFormer,Wang2023DSVT, wu2024point}, but their reliance on voxelization or fixed serialization patterns can poorly capture the irregular, complex geometry of tracking. HEPT~\citep{Miao2024HEPT} mitigates these issues by employing randomized serialization via LSH~\citep{datar2004locality}, preserving locality while reducing complexity to linear. However, its reliance on costly post-hoc clustering and the lack of physics-based evaluation limit its utility in real-time tracking.

To overcome these limitations, we present the first unified evaluation of HEPT against a strong, representative GNN pipeline~\citep{ctd2022} on realistic HL-LHC datasets with physics-grounded metrics. In addition, we introduce HEPTv2, an end-to-end efficient tracking architecture that augments the standard HEPT encoder with a lightweight transformer decoder for direct track assignment, achieving an average of 28 ms per event on the TrackML dataset~\citep{Amrouche:2021nbs,Amrouche:2019wmx}, using an NVIDIA A100. This design preserves HEPT’s efficiency while improving scalability, establishing HEPTv2 as a competitive alternative to GNN-based pipelines.

\section{Related Work}

\textbf{Traditional Methods and GNNs for Tracking.} Traditional particle tracking methods, such as the \ac{KF} and Hough Transforms, have been used extensively in high-energy physics experiments, but face performance limitations under high pile-up conditions due to their combinatorial complexity. Recent advancements in deep learning have driven the adoption of GNNs for track reconstruction, particularly through edge classification approaches~\citep{ExaTrkX:2021abe,Lazar:2022ixi,DeZoort:2021rbj}. Additionally, object condensation techniques~\citep{Lieret:2023aqg} have been explored.

\textbf{Efficient Point Transformers.} 
To avoid the quadratic complexity of standard self-attention~\citep{vaswani2017attention} when processing large point clouds, Point Transformer~\citep{Zhao2021PointTransformer} introduces a local attention mechanism that couples positional encoding with neighborhood selection.
More recently, studies have centered on point cloud serialization, which projects high-dimensional point-cloud data into one-dimensional sequences while preserving locality. Recent works such as FlatFormer \citep{Liu2023FlatFormer}, DSVT \citep{Wang2023DSVT}, and PointTransformerV3 \citep{wu2024point} adopt fixed serialization patterns, which improve tractability but struggle with irregular detector geometry. Among these, HEPT~\citep{Miao2024HEPT} stands out by employing randomized serialization via LSH, providing a theoretical guarantee of near-linear complexity and offering strong potential for tracking. There are also new ideas considering tracking as a segmentation task using Maskformers ~\citep{VanStroud:2024fau}.

\section{Methodology}

\subsection{Exa.TrkX}

We select Exa.TrkX~\citep{ctd2022} as the representative GNN-based pipeline for tracking.
Exa.TrkX consists of three stages, in which two different neural networks are employed. The first stage is the metric learning stage, where the features are pushed into an embedding space via \ac{MLP}, which then outputs higher-dimensional features, with the n-dimensional distance determining whether two hits are neighboring. A GNN is then trained to prune out false positive edges. 
Finally, the track building stage reconstructs the particle tracks from the pruned graphs, with track efficiency determined by comparing to truth. 
The Exa.TrkX pipeline is retrained and optimized to maintain high pixel tracking efficiency using the acorn framework~\citep{acorn}.

\subsection{Locality-Sensitive \ac{HEPT}}
\ac{HEPT} extends the transformer by incorporating LSH into the attention mechanism to achieve near-linear complexity while encoding a local inductive bias. Concretely, E2LSH is used with OR \& AND constructions to map points to one-dimensional buckets and apply block-diagonal attention within buckets. Each point is hashed into $m_1$ tables (OR-LSH), where each table concatenates $m_2$ independent hash functions (AND-LSH): 
$
g(x) = [h_1(x), 
\dots, h_{m_2}(x), \quad h_j(x) = \lfloor \frac{a_j \cdot x + b_j}{r} \rfloor
$. After mapping the OR \& AND hash into one single hash code, points are sorted to yield a 1D serialization, which allows the employment of block-diagonal (windowed) attention, converting irregular all-to-all interactions into regular, hardware-friendly operations.


\subsection{HEPTv2}
We extend \ac{HEPT} as an end-to-end particle tracker, mapping raw detector hits directly to track identities. The standard HEPT module is used to create encoded embeddings for each particle hit, which is then passed into a point-wise classifier that determines which points are likely to be part of a track (i.e. not background noise). On top of the encoded point features, a query-based instance decoder inspired by Mask3D~\citep{cheng2022masked, schult2022mask3d}, iteratively refines a fixed set of track hypotheses. Each learnable query attends to the selected hits, alternating self-attention, cross-attention, and feed-forward layers, outputting a per-query confidence and a dense mask indicating which hits belong to that hypothesized track. Points that were previously rejected by the classifier skip the decoding and instead have per-query mask logits computed in one shot via the learned queries.

At inference, the model converts the predicted masks into final track assignments by combining per-query confidence with per-hit mask activations and assigning each hit to its most plausible track (or to background). Small-fragment suppression removes spurious instances, yielding clean, per-event track IDs for each hit. Training is fully end-to-end with a composite loss:
$$
\mathcal{L}_{\text{total}} =
\lambda_{\text{nce}}\,\mathcal{L}_{\text{NCE}}
+ \lambda_{\text{clf}}\,\mathcal{L}_{\text{CLF}}
+ \lambda_{\text{ce}}\,\mathcal{L}_{\text{CE}}
+ \lambda_{\text{mask}}\,\mathcal{L}_{\text{BCE}}
+ \lambda_{\text{dice}}\,\mathcal{L}_{\text{Dice}}
$$
Here $\mathcal{L}_{NCE}$ is the InfoNCE \citep{oord2018representation} loss used in standard HEPT, which
enforces embedding consistency across hits for the same particle and separation from hits of different particles. $\mathcal{L}_{CLF}$ is for the per-hit binary classifier to classify points likely to be part of a track. $\mathcal{L}_{CE}$, $\mathcal{L}_{BCE}$, and $\mathcal{L}_{Dice}$ are those losses used in object segmentation in computer vision, where $\mathcal{L}_{CE}$ supervises query classification (object vs background), $\mathcal{L}_{BCE}$ provides point-wise mask accuracy, and $\mathcal{L}_{Dice}$ enforces overlap between true and predicted track-masks. To improve robustness to noise, we introduce a learning curriculum that progressively admits harder, lower-momentum hits over epochs. This allows the model to first master clean, reconstructible tracks, and then adapt to realistic backgrounds. 

\section{Experiments}

\subsection{Dataset}
The TrackML dataset~\citep{Amrouche:2021nbs, Amrouche:2019wmx} contains approximately 10,000 simulated proton-proton collision events in the \ac{LHC}, generated with Pythia 8. Each event includes a signal originating from hard QCD interactions in a top–antitop quark decay, as well as a background from 200 soft QCD interactions to realistically model pileup expected at the HL-LHC. A transverse momentum cut of $p_T >$ 150 MeV is applied for event selection. The simulated detector consists of three types of subdetector layers: pixel, short-Strip, and long-Strip. In this paper, we focus on measurements from the Pixel detector. In our study, we split 900/500/500 for train/val/test.

\subsection{Model Settings}

\textbf{Exa.TrkX.}
The GNN model used has approximately 854K parameters. Both the edge encoder network and node network have 8 layers with a hidden dimensionality of 256. The edge cut threshold is tuned to 0.1.

\textbf{HEPT/HEPTv2.} The HEPT encoder uses 3 OR hash tables, 50 regions with 2 AND hash tables, and 8 attention heads. HEPTv2 has additional 4 transformer decoder layers and 3000 instance queries that cross-attend to each hit, resulting a total of 1.1M parameters. The number of queries is set larger than the maximum track multiplicity observed across all events, ensuring full coverage.


\subsection{Pixel Tracking Performance}

\begin{figure}
\vspace{-3mm}
  \centering
  \includegraphics[width=0.49\textwidth]{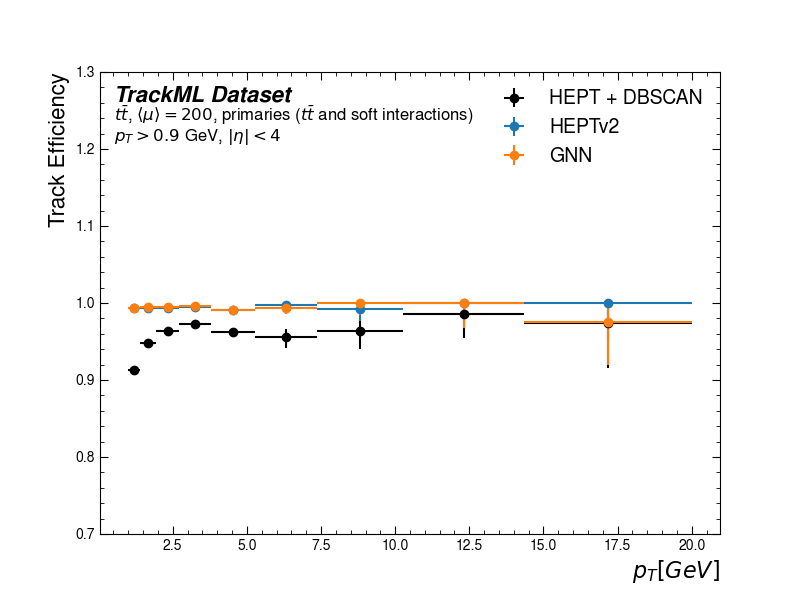}
  \includegraphics[width=0.49\textwidth]{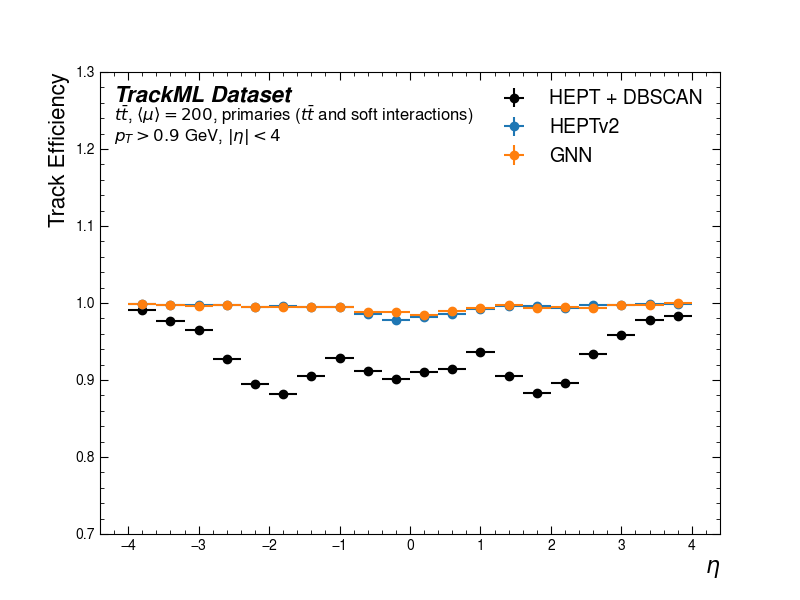}
  \vspace{-2mm}
  \caption{Tracking efficiency as a function of (a) transverse momentum, $p_{\text{T}}$, and pseudorapidity, $\eta$.}
  \label{fig:track_eff}
  \vspace{-5mm}
\end{figure}

The tracking performance of the pipeline is evaluated by the efficiency and the fake rate. In this study, charged particles are required to be within the region covered by the Pixel detectors, $|\eta| < 4.0$ and transverse momentum $p_{\text{T}} > 900~\text{MeV}$.
The matching between particles and track candidates is defined as follows: A particle is considered matched to a track candidate if at least half of the particle hits are within the candidate. A track candidate is matched to a particle if purity is higher than 50\%. For the efficiency evaluation, only particles with a minimum of three pixel hits are considered reconstructable. The tracking efficiency is defined as the fraction of matched reconstructable particles. Figure~\ref {fig:track_eff} displays the tracking efficiency as a function of particles $p_{\text{T}}$ and $\eta$ distributions. \ac{HEPT}v2 maintains competitive tracking performance compared with the Exa.TrkX pipeline across all $p_{\text{T}}$ ranges. The $\eta$ dependence is mainly due to fewer measurements in the central barrel region, which is especially impactful for the \ac{HEPT}+DBScan pipeline.

Table~\ref{tab:track_fake} presents a summary of tracking performance and fake rate. Exa.TrkX achieves a track matching efficiency of 0.994 with a fake rate of 0.002. Compared to the \ac{HEPT} + DBScan implementation, which exhibits reduced efficiency and a higher fake rate, the end-to-end \ac{HEPT}v2 pipeline demonstrates substantial improvements in both tracking efficiency and latency (also see Sec.~\ref{sec:latency}). Although its fake rate is higher than that of Exa.TrkX, this trade-off is offset by its enhanced efficiency and reduced latency, positioning  HEPTv2 as a robust and competitive approach. 

\begin{table}[htpb]
    \centering
    \begin{tabular}{c|c|c}
     Implementation & Pixel Tracking Efficiency & Fake Rate \\ \hline    
     Exa.TrkX & $\mathbf{0.994}$ & $\mathbf{0.002}$ \\
     HEPTv2    & $0.993$ & $0.113$  \\
     \ac{HEPT} + DBScan   &  $0.923$ & $0.070$  \\
    \end{tabular}
    \caption{Tracking eff. and fake rate on simulated TrackML ttbar events with an avg. pileup of 200.}
    \label{tab:track_fake}
    \vspace{-5mm}
\end{table}

\subsection{Inference Latency}\label{sec:latency}
Previous literature has shown that \ac{HEPT} encoder can be hundreds of times faster than \acp{GNN} on TrackML when run on GPUs~\citep{Miao2024HEPT}. Therefore, here we focus on comparing three HEPT configurations: the standard \ac{HEPT} encoder (which outputs only hit embeddings without track assignment), \ac{HEPT}+DBScan (implemented following~\citep{Lieret:2023aqg} on CPUs), and HEPTv2. Table~\ref{tab:extrkX_time} shows that while the standard \ac{HEPT} encoder is very fast, adding a clustering algorithm such as DBScan eliminates much of the advantage. By contrast, HEPTv2 adds only a small latency overhead while preserving \ac{HEPT}’s strong performance, highlighting its ability to deliver ultra-fast end-to-end track assignments without sacrificing tracking performance.

\begin{table}[htpb]
    \centering
    \begin{tabular}{c|c}
        Implementation & Inference Time (ms) \\ \hline
        HEPT Encoder (No track assignment) & $23.69 \pm 0.10$ \\
        HEPTv2 & $27.73 \pm 0.16$  \\
        \ac{HEPT} + DBScan & $1425.23 \pm 2.46$  \\

    \end{tabular}
    \caption{Inference time on simulated TrackML ttbar events with an avg. pileup of 200.}
    \label{tab:extrkX_time}
    \vspace{-5mm}
\end{table}
\section{Conclusion and Future Work}
This work presents a comprehensive evaluation of HEPT and HEPTv2 for particle tracking against a strong GNN baseline, using only Pixel detector measurements as a demonstration on the TrackML dataset. Our results demonstrate that HEPTv2 achieves Pixel tracking efficiency comparable to the state-of-the-art Exa.TrkX pipeline, while offering significantly reduced latency.

The HEPTv2 framework provides a promising approach for accelerating charged particle tracking with lower latency and computational overhead, making it well-suited for online trigger environments. Future work will focus on extending this architecture to incorporate full detector systems, including the Strip detector, and evaluating its impact on overall tracking performance across diverse metrics and experimental conditions.

\section*{Acknowledgments}
Y. Chou, J. Duarte, P. Li, M. Liu, S. Miao, J. P. Rodgers, and S. Hsu are supported by National Science Foundation (NSF) grant No. PHY-2117997. 
S. Miao and P. Li are also supported by NSF grant No. IIS-2239565. 
This research used resources of the National Energy Research Scientific Computing Center (NERSC), a U.S. Department of Energy Office of Science User Facility located at Lawrence Berkeley National Laboratory, operated under Contract No. DE-AC02-05CH11231.
We also acknowledge support from the NVIDIA Academic Grant Program and the IDEaS Cyberinfrastructure Awards.

\bibliography{refs}

\end{document}